# Measurement of Characteristic Impedance of Silicon Fiber Sheet based readout strips panel for RPC detector in INO

**M. K. Singh, A. Kumar, N. Marimuthu, V. Singh[*] and V. S. Subrahmanyam**

*Banaras Hindu University*
*Nuclear Physics section, Department of Physics, Vranasi-221005, India*
*E-mail*: venkaz@yahoo.com

ABSTRACT: The **I**ndia **b**ased **N**eutrino **O**bservatory (INO) is a mega science project of India, which is going to use near about 30, 000 **R**esistive **P**late **C**hambers (RPC) as active detector elements for the study of atmoshpheric neutrino oscillations. Each RPC detector will consist of two orthogonally placed readout strips panel for picking the signals generated in the gas chamber. The area of RPC detector in INO-ICAL (Iron Calorimeter) experiment will be 2m × 2m, therefore the dimension of readout strips panel will also be of 2m × 2m. To get undistorted signals pass through the readout strips panel to frontend electronics, their Characteristic Impedance should be matched with each other. For the matching of Characteristic Impedance we have used the principle of termination. In the present paper we will describe the need and search of new dielectric material for the fabrication of flame resistant, waterproof and flexible readout pickup strips panel. We will also describe the measurement of Characteristic Impedance of plastic honeycomb based readout strips panel and Silicon Fiber sheet based readout strips panel in a comparative way, and its variation under loading and with time.

[*] Corresponding author

# Contents

## 1. Introduction

The India based Neutrino Observatory is going to establish an underground laboratory in Theni district of Tamil Nadu for the study of atmospheric neutrino oscillations initially. In this laboratory the INO-ICAL detector (Iron Calorimeter) will be placed in a cavern to minimize the cosmic ray background with a minimum all round rock cover of 1 km [1]. This ICAL detector will consist of nearly 30,000 Resistive Plate Chambers which act as active detector elements. In this experiment the RPC gas chamber will be made up of either glass or/and Bakelite. These RPC detectors will be placed in between two Iron plates of thickness 56 mm with a gap of 40 mm [1]. Each of these RPC detectors will consist of two orthogonally placed (one above the RPC gas chamber and other will be below the RPC gas chamber) readout strip panel with respect to each other, in order to record particle hit coordinates in both X and Y planes [2] [3]. Therefore, near about 60,000 readout strips panels which will be used in this experimental setup to pick up the signals generated in the gas chamber due to ionization of the gas [2]. The gas composition which will be used in it are Argon, Freon and Isobutane of which Argon will be used as the target gas, Freon as an electron quencher and Isobutane will be used as photon quencher [4]. These generated signals will be transported to the further electronics using coaxial cables. In transportation of signals; one very important factor is Characteristic Impedance, which must be matched on both ends (first end is readout strips and another end is of the front-end electronic gadgets (such as pre-amplifier) connected with the cables). If this condition of impedance matching will not be perfect then there will be problem in signal transportation, such as, the signal reflected back and mixed with the original pulse, which leads to its shape distortion. Using the principle of termination, we can remove / minimise signal distortion problem. In which we add some extra resistance in between those two cables / devices and by adjusting their values we found that both will see a common value of impedance at their interface [5].

Since the whole detector (RPC gas chamber, two perpendicularly placed readout strip panels and associated preamplifiers) has to be placed in between the gaps of Iron plates [1]. Therefore instead of Characteristic Impedance matching, it is also required that the thickness of readout panels should be as minimum as possible. But this should happen without compromising the quality of signals picked up by the readout strips panel.

Though, currently plastic honeycomb is being used as a dielectric material for making the readout panel, but in perspective of the INO-ICAL underground experiment it has two major drawbacks, such as, it is not flexible and it can catch fire easily. But, for INO the required dielectric material must have the following properties, such as, it must be flame resistant or fire retarded, flexible, light weight, cost effective and locally available. Keeping these requirements we have searched several different dielectric materials and we found that it is the Silicon Fiber Sheet (SFS) which can be best suited for us, because it pertains all the above mentioned properties. Therefore, we have proceeded with it in making the readout strips panel [2] [6] [7].

We have made up Silicon Fiber Sheet (SFS) based readout strips panel in our laboratory. In making of it we have made use of BLUECOAT 9000 adhesive for attaching the Cu and Al foils, because this is the best suitable adhesive for this purpose [8]. For making the connections on the Al sheet, we found that it is difficult to do the soldering on it. To overcome this problem, we have made use of ANSOL Aluminium soldering Flux [9]. After making these connections we measured its Characteristic Impedance under different conditions.

## 2. Experimental setup and measurements

### 2.1 Impedance matching circuit

In any circuitry where we want to transmit the signal between two devices or cables, reflection of signal occurs. This must be avoided for getting proper signal. Otherwise it causes the signal to be reflected back, due to which distortion in the shape of original signal takes place. At the same time these echoes of original signal can mislead us for false results. Standard Nuclear Instrumentation (NIM) requires that all input and output devices must be of 50 ohms impedances [5]. This can be done, if we choose their Characteristic Impedances to be equal to 50 ohms. But, if on some occasion it is necessary to interconnect the two cables or devices, which have different Characteristic Impedances, then principle of termination i.e. electrical termination of a signal involves providing a terminator at the end of a wire or cable to prevent a radio frequency signal from being reflected back from the end, causing interference, is suitable for this purpose. In this termination method, we add some external impedance (either in series, parallel or in combination of both) with the impedances of two devices or cables which we want to add. By doing so, we can adjust the load seen by both of them at their interface.

### 2.2 Procedure

We have used a pulser that is designed to send the pulses through 50 ohms coaxial cables. For getting the reflected signal separately than the original pulse, we have sent the signal from coaxial cable through 10 meter long flat ribbon twisted cable of 120 ohms Characteristic

Impedance. This 120 ohms cable must be connected to a 50 ohms coaxial cable through proper terminating impedances, such that the signals coming from 50 ohms cable are matched to the 120 ohms cable, and similarly, for the reflected signals coming from the 120 ohms cable to 50 ohms cable. We have designed the circuit as shown in Figure 1 to terminate the two cables of different Characteristic Impedances. In which we have added two external resistances $R_1$ and $R_2$ in series and parallel combinations between the two cables. With this circuitry the 120 ohms cable sees a 120 ohms signal instead of 50 ohms signal and the 50 ohms cable sees a 50 ohms signal as opposed to 120 ohms signal coming from the 120 ohms cable.

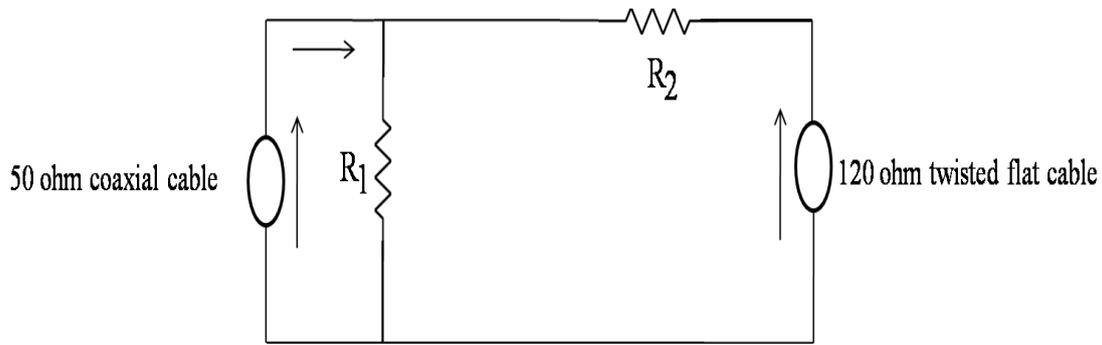

**Figure 1.** Circuit diagram of impedance matching Circuit [5].

$$R_1 \| (R_2 + 120) = 50 \qquad \text{...................Eq. 1}$$

$$R_2 + (R_1 \| 50) = 120 \qquad \text{...............Eq. 2}$$

From the circuit as shown in Figure 1, we found these two equations for terminating the two cables.

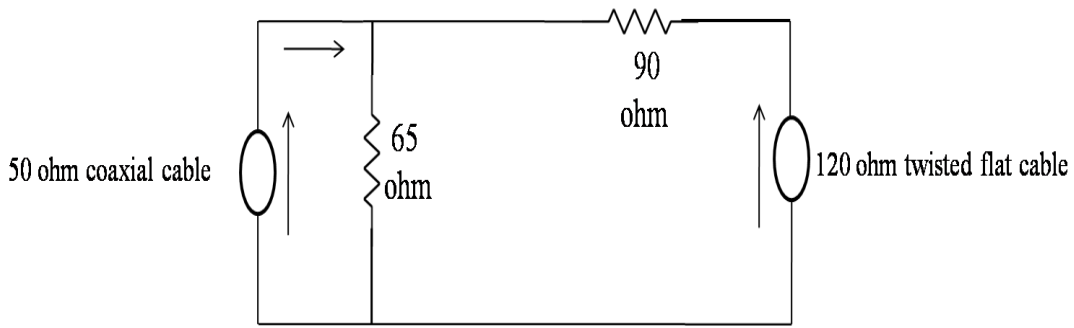

**Figure 2.** Circuit diagram of impedance matching circuit with resistance values [5].

After solving them we found that $R_1 = 65.46$ ohms and $R_2 = 91.66$ ohms. These values we have added in the circuit as shown in Figure 2, to make the impedance matching circuit. This impedance matching circuit we have added in between the 50 ohm coaxial cable and 120 ohms flat ribbon cable [5]. We have sent the pulse from pulse generator to flat ribbon cable through this impedance matching circuit.

## 2.3 Measurement of Characteristic Impedance

We measured the Characteristic Impedance of the Polycarbonate based Pickup strip panel of size 50cm × 50cm and also of Silicon Fiber Sheet (SFS) based Pickup strip panel of size 50cm × 50cm by pulse method using Scientific 20 MHz Pulse Generator SM 5035, Tektronix Digital Phosphor Oscilloscope 3054 of 500 MHz and a MARS VC97 Digital Multimeter [12] [13].

The circuit diagram with original photographs (in inset) of experimental setup is shown in Figure 3; the pulser sends a pulse through a 50 ohms coaxial cable and it feds into the terminating circuit where it passes on to a 120 ohms twisted pair flat ribbon cable without being distorted. The 120 ohms cable is about 10 meter long and it is branched to an oscilloscope for the study at about 1.5 meter from the impedance matching circuit as shown in Figure 2. One end of the 120 ohms cable is connected to the front-end of pickup strip panel. The back-end is connected with variable resistance for which we have used the BOURNS 3296-3/8" trimpot square trimming potentiometer (COSTA RICA) as shown in inset of Figure 3, along with multimeter to measure the change in resistance [13]. We have designed and tested many pickup strip panels using various dielectric materials but could not achieve the desired results. We obtained the desired results only with Silicon fiber sheet dielectric material.

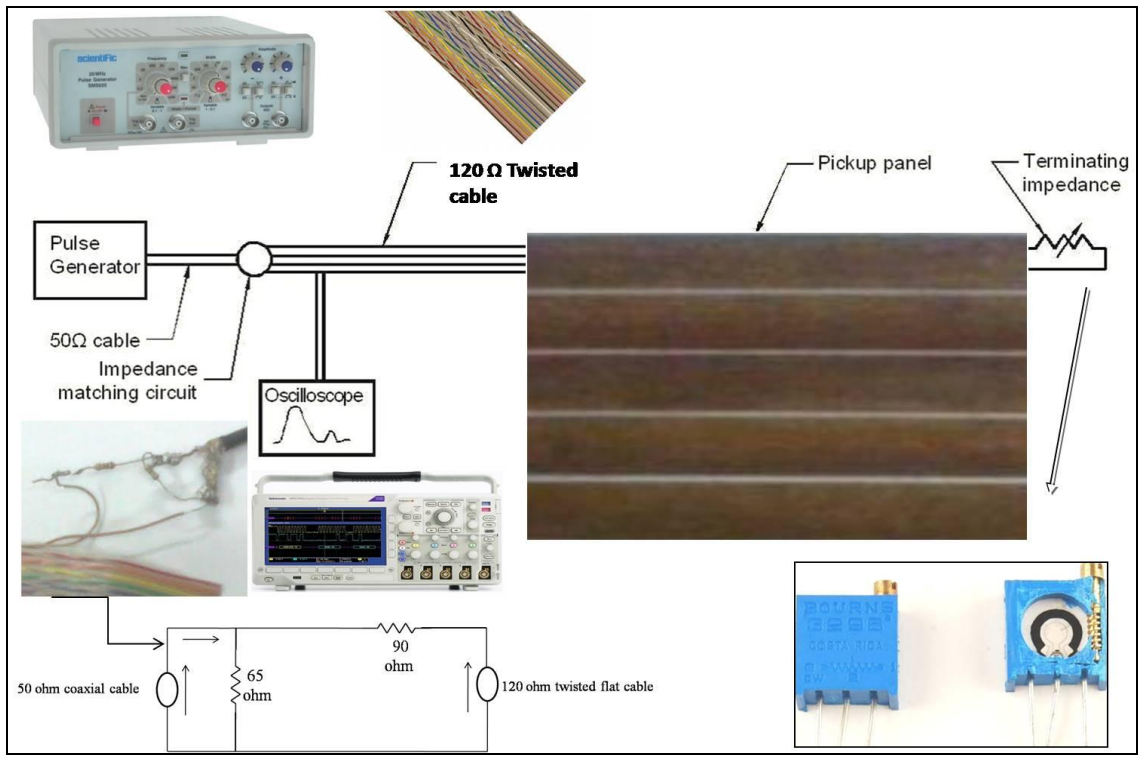

**Figure 3.** Experimental arrangements for the measurement of Characteristic Impedance [12].

## 2.4 Study of Pulse Analysis

With this experimental arrangement we want to determine the following valuable information which will leads us to make a good quality readout strips panel.

**1)** The thickness of dielectric material for which we will get the minimum signal reflection.

– 4 –

**2)** The value of corresponding terminating impedance at the backend of the pickup strip panel.

Figure 4 shows the original and reflected pulse from one of the strips of the pickup panel with open end circuit and similarly we have also observed for each strip. The pulse analysis has been tabulated below in Table 1. We observed that, the maximum fluctuation of noise from the base or pedestal line, between input pulse and frontend reflected pulse is $5\sigma$ ($1\sigma = 40$ mV) i.e. 200 mV.

**Table 1:** Signal pulse analysis without termination of far end.

|  | **Input Pulse** | **Frontend Reflection** | **Backend Reflection** |
|---|---|---|---|
| FWHM (ns) | 9 | 4 | 10 |
| Rise time (ns) | 10 | 3 | 8 |
| Fall time (ns) | 16 | 4 | 18 |
| Amplitude (mV) | 800 | 80 | 320 |
| Time delay (ns) | - | 81 | 92 |

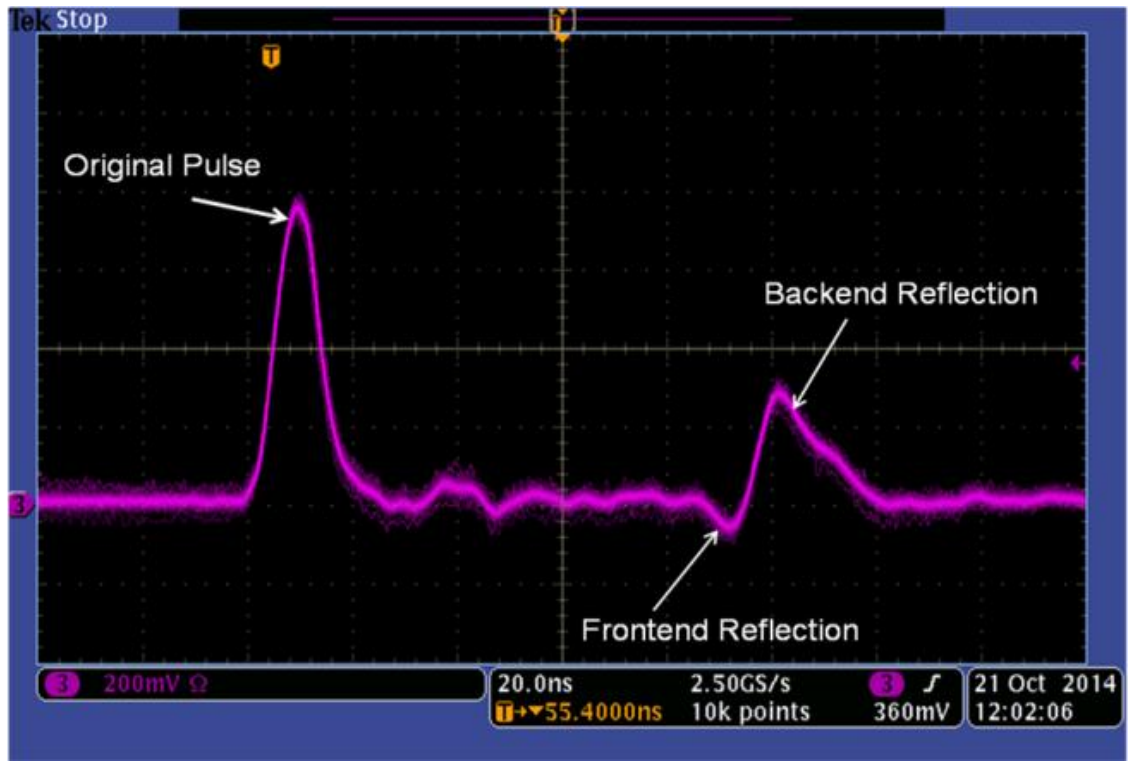

**Figure 4.** Original signal of pulse generator (Left) along with its reflected signal (Right) from the far open end [12].

For each strip we sent a pulse, and monitored the pulse reflected back from both front-end and back-end. The back-end reflections have more chances to interfere with the original signal and therefore it must be minimized as much as possible. For eliminating the back-end reflection we connected a terminating resistor at the back-end of strips, of value equal to the Characteristic Impedance of the strip. For this purpose we used a multi-turn trimming potentiometer and varied the resistance until the back-end reflection had been minimized. The value of resistance at the minimized reflection is the Characteristic Impedance of that strip that has been measured by multimeter after disconnecting multi-turn trimming potentiometer from the circuit. Similarly we have measured the Characteristic Impedance value of every strip in the readout panel.

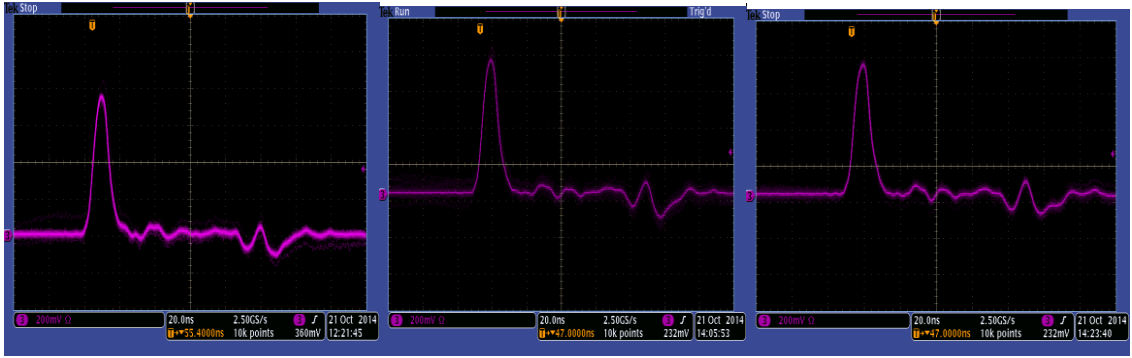

| (a) | (b) | (c) |

**Figure 5.** Oscilloscope snapshots for minimization of reflected pulse [12].

A similar procedure has been followed for the study of plastic honeycomb based pickup strips panel. In Figure 5, we have shown the snapshots of oscilloscope display from left to right for Silicon Fiber Sheet (thickness 5.00 mm), plastic honeycomb (thickness 4.53 mm) and Silicon Fiber Sheet (thickness 4.00 mm) based readout panels respectively, for optimum minimization of back-end reflection. This indicates that 5.00 mm thick Silicon Fiber Sheet based readout strips panel will have minimum reflection for the same applied terminating resistance which is 48.5 ohms.

We measured the thickness of pedestal (before the rise of the original pulse) equal to 40 mV which is equivalent to $1\sigma$. In a similar way, we measured fluctuation in the region between after the fall of original pulse and before the front-end reflection in terms of sigma which is equal to $3\sigma$, $3\sigma$ and $3\sigma$ for Figure 5(a), (b) and (c), respectively, and also measured the back-end reflection pulse amplitude which is equal to $4\sigma$, $5\sigma$ and $5\sigma$ for Figure 5(a), (b) and (c), respectively, which are also tabulated in Table 2.

We measured from these snapshots as shown in Figure 5 that the back-end reflection is minimum in case of 5.00 mm thick Silicon Fiber Sheet based readout strips panel in comparison with other combinations such as the plastic honeycomb of thickness 4.53 mm and 4.00 mm thick Silicon Fiber Sheet based panels. Hence we may conclude that the Silicon Fiber Sheet of thickness 5.00 mm is most suitable for making readout strip panel for RPC detector. Some valuable information related to the front-end and back-end reflections, and the fluctuation of noise is listed in Table 2. From this table one can see that variation of terminating impedance have no effect on frontend reflection's amplitude and only affects on back-end reflection.

**Table 2:** Fluctuation in noise and effect on front-end and back-end reflections.

| Readout Strips Panel | Maximum fluctuation of noise from base line ($1\sigma$ = 40 mV) | Effect on frontend reflection ($1\sigma$ = 40 mv) | Effect on backend reflection ($1\sigma$ = 40 mv) |
|---|---|---|---|
| Silicon Fiber Sheet (5.00 mm) | $5\sigma$ | No effect observed and its amplitude remain at $3\sigma$ | $4\sigma$ |
| Plastic Honeycomb (4.53 mm) | $5\sigma$ | No effect observed and its amplitude remain at $3\sigma$ | $5\sigma$ |
| Silicon Fiber Sheet (4.00 mm) | $5\sigma$ | No effect observed and its amplitude remain at $3\sigma$ | $5\sigma$ |

**2.5 Variation of Characteristic Impedance with the thickness of readout strips panel**

To study the behaviour of Characteristic Impedance with the thickness of pickup panel, we have made single readout strip panel of thickness 1.00 mm, 2.00 mm, 3.00 mm, 4.00 mm and 5.00 mm using Silicon Fiber Sheet as dielectric material. After that following the above procedure of measurement of Characteristic Impedance as discussed in section 2.3, we have measured the Characteristic Impedance of each pickup strip panel and the results are tabulated in Table 3. We would like to mention here that during all measurements we have kept the relative humidity and temperature of our laboratory was 43% and 25 $^0$C, respectively.

**Table 3.** Results of Characteristic Impedance measurement are listed [12].

| S. No. | Thickness of Pickup strip panel (mm) | Strip width/Thickness, Ratio | Terminating impedance (ohm) |
|---|---|---|---|
| **1.** | 1.00 | 28.0 | 134.0 |
| **2.** | 2.00 | 14.0 | 116.7 |
| **3.** | 3.00 | 9.3 | 93.7 |
| **4.** | 4.00 | 7.0 | 73.0 |
| **5.** | 5.00 | 5.6 | 50.0 |

The thickness of the metallic foil used in making the readout strips is mentioned below in Table 4. The strip width has been kept constant at 28.0 mm during Characteristic Impedance measurement. Therefore, as the thickness of readout strip panel is increased, the ratio of strip width to thickness decreases as shown in Table 3.

**Table 4.** Thickness of metallic foils used in making readout strips panel.

| S. No. | Thickness → | Pickup strip panel (Polycarbonate) | Pickup strip panel (Ceramic Foam) |
|---|---|---|---|
| **1.** | **Copper** | 0.09 mm | 0.04 mm |
| **2.** | **Aluminum** | 0.17 mm | 0.01 mm |

From Table 3, we have plotted a graph to find out the variation of Characteristic Impedance with the thickness of dielectric material, which has been shown in Figure 6. From this plot one can see that it is the 5.00 mm thickness of readout strips panel which is required to minimize the reflected signal, as we have also observed this in our earlier section 2.4 and obtain the impedance of 50 Ω which is almost same to the Characteristic Impedance of cables and connectors. Therefore, we may prefer to make the readout strips panel of thickness 5.00 mm. This will not only remove the problem of mismatch of Characteristic Impedances between readout panel and frontend electronics but also fulfill the requirement of minimum possible strip panel thickness because, we have only a gap of 40.00 mm between the Irons plates of ICAL detector for sandwiching the RPC detector system [1].

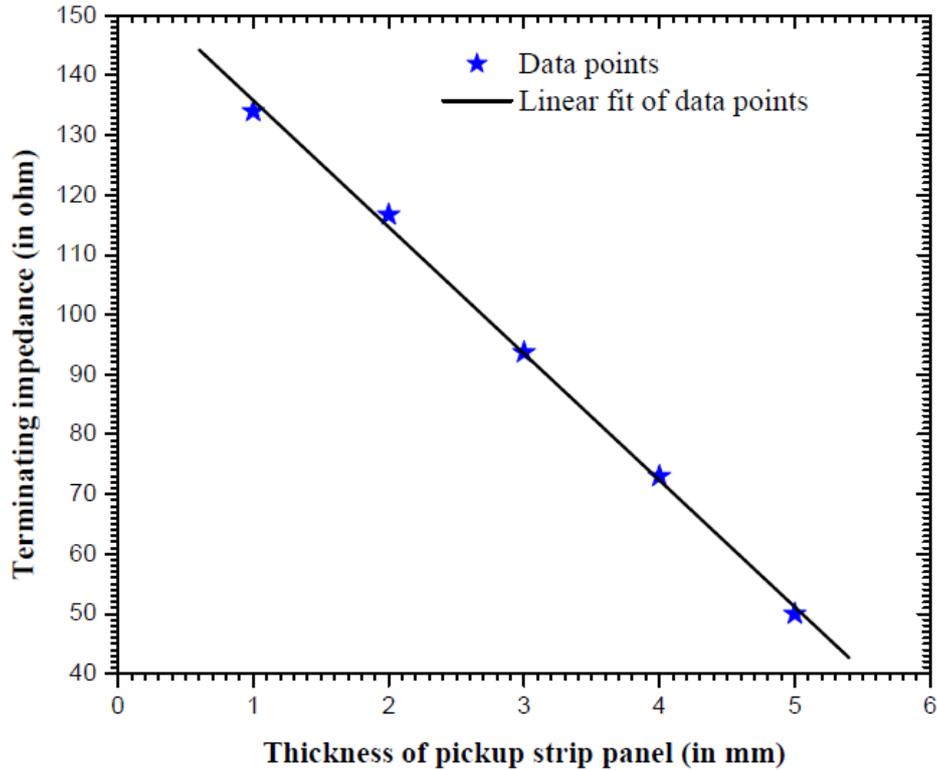

**Figure 6.** Variation of Characteristic Impedance with thickness of single Silicon Fiber Sheet based readout strip panels [12].

## 3. Variation of Characteristic Impedance with pressure and time

Now-a-days any experimental setup must be last for longer time to take consistent data and INO – ICAL will also be one of such kind of experiment, therefore it is important to check the variation in Characteristic Impedance with time.

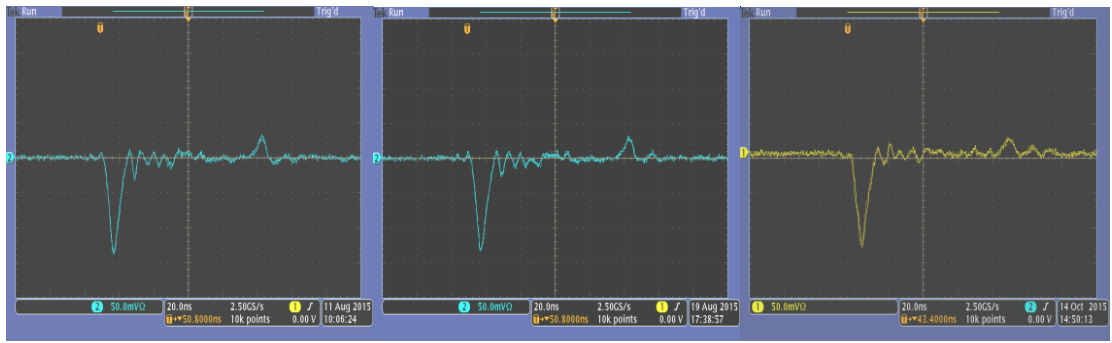

**Figure 7.** Snapshots of oscilloscope display showing the variation in reflected signal with time (after putting the weights). Dates of snapshots taken are as follows from left to right: 11 Aug 2015, 19 Aug 2015, and 14 Oct 2015 in a continuous manner.

Another important issue is that since the Silicon Fiber Sheet is flexible and little softer than plastic honeycomb, therefore, it is also important to check the variation in Characteristic Impedance of the readout strips panel with the weight of glass RPC gas chamber on it.

To study these two issues we have prepared an experimental setup in which we have pressed the single strip Silicon Fiber Sheet based readout pickup panel with more than 30 times weight of glass RPC gas chamber using Lead (Pb) bricks and then using the circuit as shown in Figure 3, we measured and monitored the Characteristic Impedance up to 72 days continuously from August 11, 2015 to October 14, 2015. The oscilloscope snapshots are shown in Figure 7 in which we can see the variation in the reflected signal after terminating the single Silicon Fiber Sheet readout strip panel's far end by 50 ohms with the help of BOURNS 3296-3/8" trim-pot square trimming potentiometer (COSTA RICA) [13]. The maximum reflection amplitude was approximately 3σ (1σ = 10 mV) which was sustained up to the final date of measurement. For better understanding of change in the back-end reflected pulse amplitude, we have taken the oscilloscope snapshot of reflected signal with input pulse without weight application, and it is shown in Figure 8. Figure 8 also shows that the maximum reflection retained before the application of weight was approximately 3σ (1σ = 10 mV).

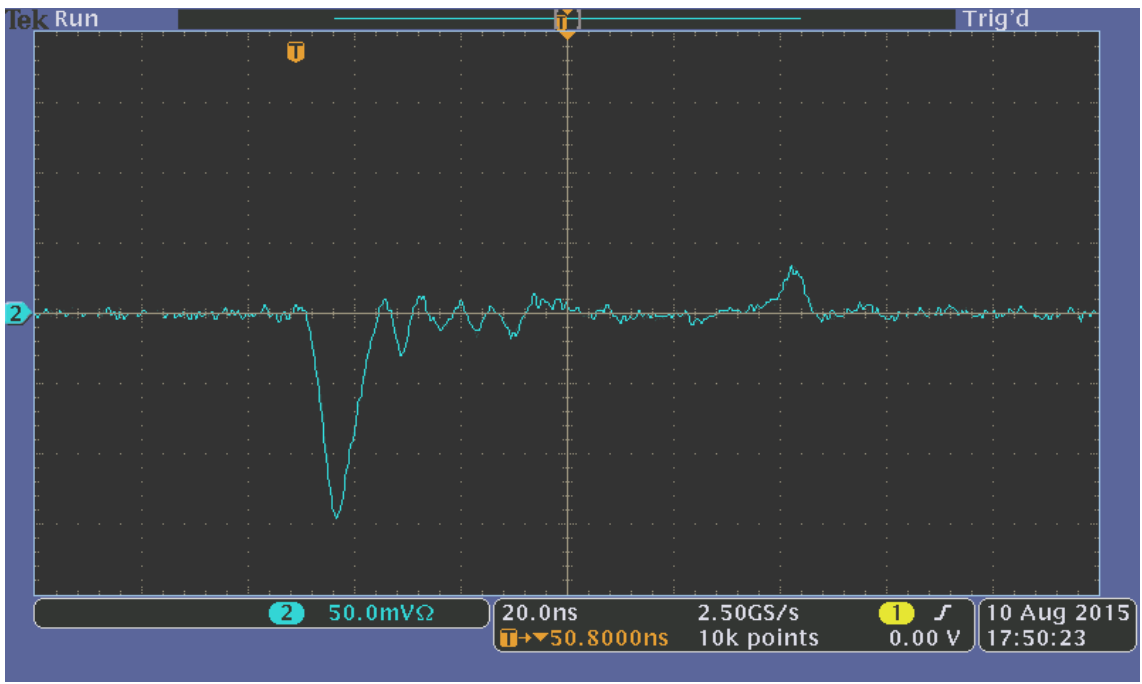

**Figure 8.** Oscilloscope snapshot taken before putting weights on the single strip readout pickup panel.

Looking at these three as shown in Figure 7 and many more snapshots which we have taken one on every day basis we conclude that there is no variation in the signal reflection with time, and as far as the variation in Characteristic Impedance with the weight of glass RPC gas chamber is concerned, there also seems to be no significant change. We observed that there is no change in the reflection even under the effect of almost 30 times weight of RPC gas chamber is applied.

Figure 9, shows the variation in amplitude of reflected pulse with time under the application of weights. It can be seen from Figure 9 that there is almost no variation in reflected signal amplitude.

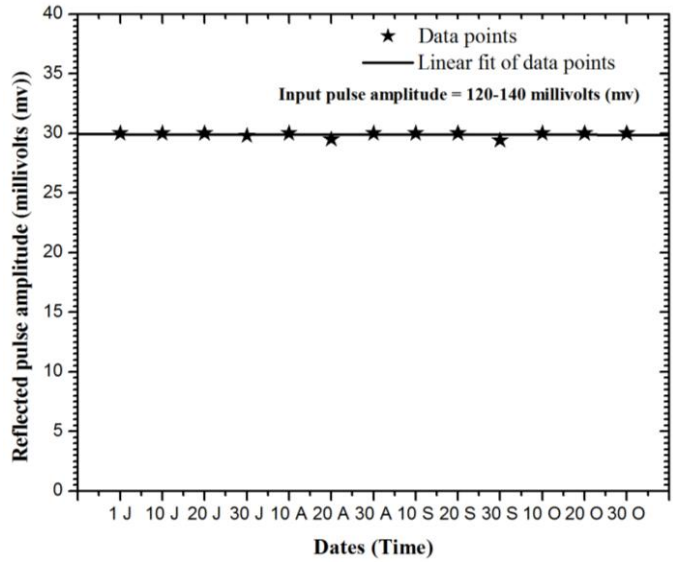

**Figure 9.** Variation of Characteristic Impedance with time.

## 4. Variation in Characteristic Impedance of each strip in readout strips panel

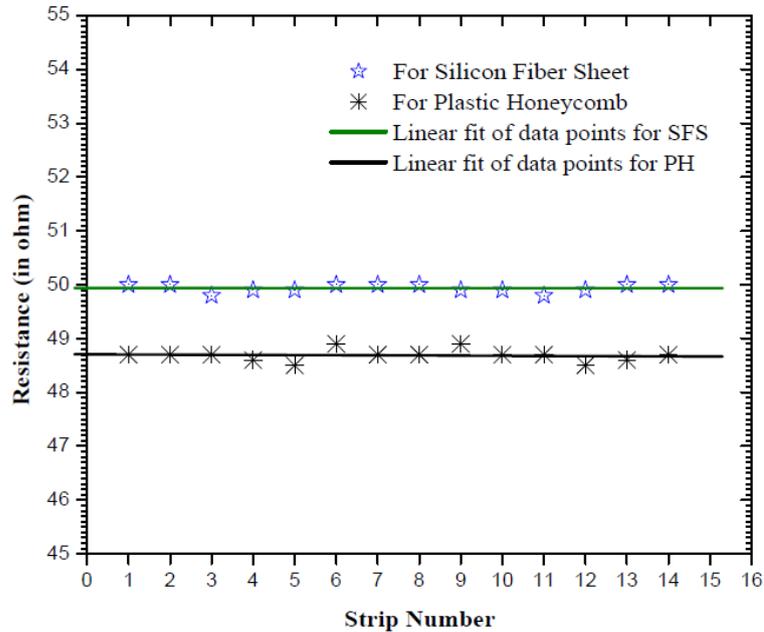

**Figure 10.** Characteristic Impedance of each strip in readout pickup panel.

Finally, we have measured the Characteristic Impedance of each strip of the readout pickup panel to know if there is any significant variation in the Characteristic Impedance of each strip due to any effects of neighboring strips. Following the same procedure as mentioned in section 2.3, we have measured it, and found that there is no significant variation in the Characteristic Impedance of any strips of the Silicon Fiber Sheet based readout pickup strips panel as shown in Figure 10. It also shows that slight variation in Characteristic Impedance is almost similar as the Plastic honeycomb based readout strips panel's strip. From Figure 10, one can observe that for each strip the Characteristic Impedance value is almost same as the other strips and lying at 49.9 ohm and 48.7 ohm for Silicon Fiber Sheet and Polycarbonate based pickup strips panels, respectively. Therefore, we may conclude that Silicon Fiber Sheet based readout pickup strips panel is working properly and can be used for picking up the signals from Resistive Plate Chamber Detector.

## 5. Results and conclusion

As we know that Silicon Fiber Sheet is mechanically flexible, thermally flame resistant and chemically water proof i.e. moisture or relative humidity has no effect on it. All these properties are required for the materials used in the underground laboratories. We observed that 5 mm thick Silicon Fiber Sheet based readout strips panel has minimum signal reflection, in comparison to the 1 mm, 2 mm, 3 mm and 4 mm, with Characteristic Impedance value is 49.9 ohms (a pitch of 28 mm, is kept constant). Therefore, in order to get the least signal reflection for performing the experiment with RPC detector in INO, we propose the use of Silicon Fiber Sheet based readout strips panel of thickness 5 mm will be more suitable. We also observed that the Characteristic Impedance of Silicon Fiber Sheet based readout strips panel decreases with the increase in thickness. There is no variation in the signal reflection we observed starting from day one to day seventy two, continuously, which indicates the Characteristic Impedance of Silicon Fiber Sheet based readout strips panel, has no variation in its Characteristic Impedance even after a long use. On the basis of our study on Silicon Fiber Sheet based readout strips panel in a comparative way with polycarbonate i.e. plastic honeycomb based read out strips panel, we may conclude that Silicon Fiber Sheet Based readout strips panel has similar performance to the plastic honeycomb based readout strips panel with less signal reflection and also its Characteristic Impedance is more close to 50 ohms than plastic honeycomb based readout strips panel with a Characteristic Impedance of 48.7 ohms.

## Acknowledgments

Authors are grateful to the Department of Science and Technology (DST), New Delhi, India for providing financial support and India-based Neutrino Observatory (INO) Collaboration for their continuous support and fruitful discussions. Authors are deeply thankful to Shri R. R. Shinde of TIFR, Mumbai for his valuable suggestions and proper guidance, which helped in performing the present experimental works.